\documentstyle[12pt]{article}
\begin{document}
\thispagestyle{empty}
\begin{center}
\Large{ $GL_q(N)$-COVARIANT BRAIDED\\
DIFFERENTIAL BIALGEBRAS}
\end{center}

\vspace{.5cm}

\begin{center}
\large{A.P.\,ISAEV}${}^{\,*}$ \large{and
A.A.VLADIMIROV}${}^{\,\diamond}$
\end{center}
\begin{center}
\large{Bogolubov Theoretical Laboratory, \\
Joint Institute for Nuclear Research, \\
Dubna, Moscow region 141980, Russia}
\end{center}

\vspace{1cm}

\begin{center}
ABSTRACT
\end{center}

We study a possibility to define the (braided) comultiplication
for the $GL_q(N)$-covariant differential complexes on some quantum
spaces. We discover such {\em differential bialgebras} (and Hopf
algebras) on the bosonic and fermionic quantum hyperplanes (with
additive coproduct) and on the braided matrix algebra $BM_q(N)$ with
both multiplicative and additive coproducts. The latter case is
related (for $N=2$) to the $q$-Minkowski space and $q$-Poincare algebra.

\vspace{4cm}

${}^*$ E-mail: isaevap@theor.jinrc.dubna.su

\vspace{.3cm}

${}^\diamond$ E-mail: alvladim@thsun1.jinr.dubna.su

\pagebreak

{\bf 1.} Throughout the recent development of differential calculus
on quantum groups and quantum spaces, two principal and closely related
concepts are readily seen. One of them, initiated by
Woronowicz~\cite{Wo}, is known as {\em bicovariant differential
calculus} on quantum groups. Its characteristic feature is the
covariance under the left and right ``group shifts'' $\Delta_L$ and
$\Delta_R$ acting upon the differential complex in a consistent way.
Brzezinski~\cite{Br} has shown that this corresponds to existence of a
{\em differential bialgebra}, i.e. a bialgebra structure with honest
coproduct $\Delta$ on the whole algebra of coordinate
functions and their differentials. This allows one to treat all the
subject using the standard Hopf-algebra technique.

Another concept, introduced by Wess and Zumino~\cite{WZ}
(see also \cite{Man}),
proceeds from the requirement of covariance of the differential complex
on a quantum space with respect to the coaction of
some outer quantum group considered as a group of symmetry.
In other words, the corresponding differential algebra
must be a covariant comodule as well.

In the present letter, we want to unify both concepts by formulating
the following set of conditions to be satisfied by $q$-deformed
differential calculi:

$\alpha$) associative algebra of generators and differential forms
is respected by the (co)action of some quantum group;

$\beta $) external differentiation $d$ obeys $d^2=0$ and the usual
(graded) Leibnitz rule;

$\gamma $) differential algebra admits a (braided) coproduct of the
form~\cite{Br}
\begin{equation}
\Delta(a)=a_{(1)}\otimes a_{(2)}\,,\ \ \ \ \Delta(da)=
da_{(1)}\otimes a_{(2)}+a_{(1)}\otimes da_{(2)}\,. \label{0}
\end{equation}
Following these criteria, we obtain several examples of
$GL_q(N)$-covariant differential bialgebras: the braided matrix algebra
$BM_q(N)$ (with additive and multiplicative coproducts)  and
also bosonic and fermionic quantum hyperplanes with additive
coproduct. The first example seems to be of special interest because
$BM_q(2)$ is presently considered as a candidate to the role of the
$q$-Minkowski space~\cite{C-WSSW}-\cite{MM}.

\vspace{.3cm}

{\bf 2.} To formulate and study the quantum-group-covariant
differential calculus, the $R$-matrix formalism~\cite{FRT} proved to
be extremely convenient. Let us first consider the case of braided
matrix algebra $BM_q(N)$ with the generators $\{1,u_j^i\}$ (the latter
form the $N\!\times\! N$-matrix $u$) and relations \begin{equation}
R_{21}u_2R_{12}u_1=u_1R_{21}u_2R_{12}\,,  \label{1}
\end{equation}
where $R$ is the $GL_q(N)$ $R$-matrix~\cite{FRT,Ji}. The multiplication
rule (\ref{1}) is invariant under adjoint coaction of $GL_q(N)$,
\begin{equation}
u^i_j\rightarrow T^i_mS(T^n_j)\otimes u^m_n\,, \ \ {\rm or}\ \
u\rightarrow TuT^{-1}\,, \label{2}
\end{equation}
where $T^i_j$ obey the relations
\begin{equation}
R_{12}T_1T_2=T_2T_1R_{12} \label{3}
\end{equation}
and commute with $u^m_n$\,. Eq.(\ref{1}) (``reflection equations'')
first appeared in the course of investigations of 2-dimensional
integrable models on a half-line (see~\cite{KS} and references
therein). Further it was studied by Majid~\cite{Maj-ex} within the
general framework of braided algebras.

 From now on we prefer to use the notation $a\otimes 1\equiv a,
1\otimes a\equiv a'$ for any element $a$.
The matrix notation will also be slightly modified \cite{IsPya}
to simplify the relevant calculations:
$$
P_{12}R_{12}\equiv\hat{R}_{12}\equiv R\,, \ R^{-1}\equiv\overline{R}\,,
 \ u_1\equiv u\,, \ u_1'\equiv u'\,.
$$
Thus, the Hecke condition for $R$ reads
\begin{equation}
R-\overline{R}=q-q^{-1}\equiv\lambda \,, \label{4}
\end{equation}
whereas (\ref{1}) becomes simply
\begin{equation}
RuRu=uRuR\,. \label{5}
\end{equation}

Differential complex on $BM_q(N)$ is defined by (\ref{5}) and
\begin{equation}
R\,u\,R\,du=du\,R\,u\,\overline{R}\,, \label{6}
\end{equation}
\begin{equation}
R\,du\,R\,du=-du\,R\,du\,\overline{R}  \label{7}
\end{equation}
(here and below we omit the wedge product symbol $\wedge$ in the
multiplication of differential forms). Of course, one could
perfectly well use $$ \overline{R}\,u\,R\,du=du\,R\,u\,R $$
instead of (\ref{6}): these possibilities are absolutely parallel. We
should also note that an agreement of (\ref{6}) with (\ref{5}) (via the
Leibnitz rule), and some other formulas below, rely heavily on the
Hecke condition (\ref{4}) that is specific to the $GL_q(N)$ case.

Commutational relations (\ref{6}),(\ref{7}) have been found in the
component form for $N=2$ ~\cite{OSWZ} in the context of the
$q$-Poincare algebra, and then recast into the $R$-matrix form
in~\cite{AKR}.  Besides that, eq.(\ref{5}) is known~\cite{Maj-ex} to
admit the multiplicative coproduct \begin{equation}
\Delta(u^i_j)=u^i_k\otimes u^k_j\,, \ \ \ {\rm or} \ \ \Delta(u)=
u\otimes u\equiv u\,u'\,, \label{8} \end{equation} provided the
nontrivial braiding relations \begin{equation}
\overline{R}\,u'\,R\,u=u\,\overline{R}\,u'\,R \label{9}
\end{equation}
are used for commuting primed $u$-matrices with unprimed ones.
Recall
{}~\cite{Maj-kit} that the braiding transformation $\Psi : A\otimes
B\rightarrow B\otimes A$, where $A$ and $B$ are covariant comodules of
a quantum group, is a map which commutes with the group coaction and,
therefore, produces a covariant recipe for multiplying tensor products
of generators:
$$ (1\otimes a)\,(b\otimes 1)\equiv a'\,b=\Psi(a\otimes b)\,. $$
For instance, eq.(\ref{9}) is induced by the corresponding universal
${\cal R}$-matrix through the (somewhat symbolic) relation
$$ \Psi(u'\otimes u)=<TuT^{-1}\otimes Tu'T^{-1}\,,{\cal R}>\,. $$

Now let us examine whether a map of the form (\ref{0}) (see also
{}~\cite{IP,IsPop}),
\begin{equation}
\Delta(du)=du\otimes u+u\otimes du\equiv du\,u'+u\,du'\,, \label{10}
\end{equation}
together with (\ref{8}) yields a proper coproduct for the whole
algebra (\ref{5})-(\ref{7}). Our statement is that it really does.
Moreover, two different sets of the braiding relations can be used here
equally well: one based on (\ref{9}),
\begin{equation}
\label{11}
 \;\;\;\;\;  \left\{
\begin{array}{l}
 \overline{R}\,u'\,R\,u=u\,\overline{R}\,u'\,R\,, \\
 \overline{R}\,du'\,R\,u=u\,\overline{R}\,du'\,R\,, \\
 \overline{R}\,u'\,R\,du=du\,\overline{R}\,u'\,R\,,   \\
 \overline{R}\,du'\,R\,du=-du\,\overline{R}\,du'\,R\,,
\end{array}
\right.
\end{equation}
and the other,
\begin{equation}
\label{12}
 \left\{
\begin{array}{l}
 \overline{R}\,u'\,R\,u=u\,R\,u'\,R\,, \\
 \overline{R}\,du'\,R\,u=u\,R\,du'\,R\,, \\
 \overline{R}\,u'\,R\,du=du\,R\,u'\,R\,, \\
 \overline{R}\,du'\,R\,du=-du\,R\,du'\,R\,.
\end{array}
\right.
\end{equation}

The proof that (\ref{10}) is an algebra homomorphism
is straightforward. For illustration, we
explicitly verify one of the required conditions using, say, the
braiding (\ref{12}):
$$ R\,\Delta(u)\,R\,\Delta(du)=R\,u\,\underline{u'\,R\,du}\,u'+
R\,u\,\underline{u'\,R\,u}\,du' $$
$$=\underline{R\,u\,R\,du}\,R\,\underline{u'\,R\,u'}+
\underline{R\,u\,R\,u}\,\underline{R\,u'\,R\,du'}=
du\,R\,\underline{u\,\overline{R}\,R\,R\,u'\,R}\,u'\,\overline{R} $$
$$ +u\,R\,\underline{u\,R\,du'\,R}\,u'\,\overline{R}=
du\,R\,\overline{R}\,u'\,R\,u\,u'\,\overline{R}+
u\,R\,\overline{R}\,du'\,R\,u\,u'\,\overline{R} $$
$$ =(du\,u'+u\,du')\,R\,u\,u'\,\overline{R}=
\Delta(du)\,R\,\Delta(u)\,\overline{R} $$
(underlining indicates the parts to which the next operation is
to be applied). Similar calculations for eq.(\ref{7}) are in fact
optional, because their result can be foreseen by differentiating the
equality just obtained. Finally, we stress that the coassociativity
of (\ref{8}) and (\ref{10}) is evident.

Note that the first equation in (\ref{12})
has already been used as a
braiding in~\cite{Mey} to make the algebra (\ref{5}) a bialgebra with
{\em additive} coproduct (see below).

For both versions of the braiding relations, (\ref{11}) and (\ref{12}),
the differential complex (\ref{5})-(\ref{7}) admits the coproduct
(\ref{8}),(\ref{10}); so $BM_q(N)$ becomes a differential bialgebra.
A counit is defined in an obvious way,
$$ \varepsilon (1)=1\,, \ \ \varepsilon (u)={\bf 1}\,, \ \ \
\varepsilon (du)=0\,. $$
Moreover,
the braided antipode can also be introduced in complete analogy with
the differential $GL_q(N)$ case \cite{IsPop},
thus making $BM_q(N)$ a differential Hopf algebra.

\vspace{.3cm}

{\bf 3.} Now we proceed to another, additive, algebra map
\begin{equation}
\Delta(u)=u\otimes 1+1\otimes u\equiv u+u'\,, \label{13}
\end{equation}
\begin{equation}
\Delta(du)=du\otimes 1+1\otimes du\equiv du+du' \label{14}
\end{equation}
which proves to be a second coproduct on $BM_q(N)$. It has been found
in~\cite{Mey} that (\ref{13})  is compatible with (\ref{5}),
provided the braiding is defined by
the first line in (\ref{12}). Our
result is that the whole differential complex (\ref{5})-(\ref{7})
admits (\ref{13}),(\ref{14}) as a coproduct if we use one of the
following four sets of the braiding relations:
\begin{equation}
\label{15}
 \;\;\;\;\;  \left\{
\begin{array}{l}
 R\,u'\,R\,u=u\,R\,u'\,\overline{R}\,, \\
 R\,u'\,R\,du=du\,R\,u'\,\overline{R}-\lambda\,u\,R\,du'\,, \\
 R\,du'\,R\,u=u\,R\,du'\,R\,,  \\
 R\,du'\,R\,du=-du\,R\,du'\,R\,;
\end{array}
\right.
\end{equation}

\vspace{.3cm}

\begin{equation}
\label{16}
 \;\;\;\;\; \left\{
\begin{array}{l}
 \overline{R}\,u'\,R\,u=u\,R\,u'\,R\,, \\
R\,u'\,R\,du=du\,R\,u'\,R\,,             \\
 du'\,R\,u\,\overline{R}=R\,u\,R\,du'+\lambda \,du\,R\,u'\,, \\
 R\,du'\,R\,du=-du\,R\,du'\,R\,;
\end{array}
\right.
\end{equation}
the remaining two sets are obtained from (\ref{15}) and
(\ref{16}) by changing the position of the prime $u \leftrightarrow u'$
(it corresponds to the
inverse braiding transformation $\Psi^{-1}$).

In this case, it is also easy to define a counit $\varepsilon $ and an
antipode $S$,
$$ \varepsilon (u)=\varepsilon (du)=0\,, \ \ \ S(u)=-u\,,\ \ \
S(du)=-du\,, $$
thus completing the construction of the differential Hopf algebra (with
additive coproduct) on $BM_q(N)$.

\vspace{.3cm}

{\bf 4.} As it has been pointed out in~\cite{AKR}, the braided matrix
algebra $BM_q(2)$ can also be interpreted as a quantum hyperplane
($q$-Minkowski space) for
the quantum Lorentz group $SO_q(3,1)$. The coordinate algebra of
a quantum hyperplane is known to admit, in a quite general situation,
an additive bialgebra structure~\cite{Maj-add}. So, a natural question
arises: can one define (additive) differential bialgebras on the
hyperplanes related to arbitrary Yang-Baxter $R$-matrices? We can answer
this question affirmatively for $R$-matrices of the Hecke type
(\ref{4}), in particular, for the $GL_q(N)$-covariant differential
complexes proposed by Wess and Zumino~\cite{WZ}:
$$ R\,x_1\,x_2=c\,x_1\,x_2\,, $$
\begin{equation}
c\,R\,dx_1\,x_2=x_1\,dx_2\,, \label{17}
\end{equation}
$$ c\,R\,dx_1\,dx_2=-dx_1\,dx_2 $$
($c$ is equal to $q$ for the bosonic and to $-q^{-1}$ for the fermionic
hyperplanes). These commutation relations are invariant under the
coaction of $GL_q(N)$
\begin{equation}
x^i\rightarrow T^i_j\otimes x^j\,,\ dx^i\rightarrow T^i_j\otimes dx^j\,
,\ \  {\rm or}\ \ \ x\rightarrow T\,x\,,\ dx\rightarrow T\,dx \label{18}
\end{equation}
(see~\cite{Su} for the generalization to the case $dx\rightarrow
T\,dx+dT\,x$), and admit the differential Hopf algebra structure with
the coproduct
\begin{equation}
\Delta(x)=x+x'\,, \ \ \ \ \Delta(dx)=dx+dx' \label{19}
\end{equation}
and the counit and antipode given by
$$ \varepsilon (x)=\varepsilon (dx)=0\,,\ \ S(x)=-x\,,\ \ S(dx)=-dx\,,
$$ if one of the following four sets of the braiding relations is
implied:
\begin{equation}
\label{20}
 \;\;\;\;\; \left\{
\begin{array}{l}
 R\,x_1'\,x_2=c^{-1}\,x_1\,x_2'\,,  \\
 R\,x_1'\,dx_2=c\,dx_1\,x_2'\,,    \\
 R\,dx_1'\,x_2=c^{-1}\,x_1\,dx_2'-\lambda\,dx_1\,x_2'\,,  \\
 R\,dx_1'\,dx_2=-c\,dx_1\,dx_2'\,; \\
\end{array}
\right.
\end{equation}

\vspace{.3cm}

\begin{equation}
\label{21}
 \;\;\;\;\; \left\{
\begin{array}{l}
 \overline{R}\,x_1'\,x_2=c\,x_1\,x_2'\,, \\
c^{-1}\,x_1'\,dx_2=R\,dx_1\,x_2'+\lambda\,x_1\,dx_2'\,,  \\
 R\,dx_1'\,x_2=c\,x_1\,dx_2'\,, \\
 R\,dx_1'\,dx_2=-c\,dx_1\,dx_2'\, . \\
\end{array}
\right.
\end{equation}
Two other sets can be obtained from these relations
by substitution $x \leftrightarrow x'$ and
correspond to the inverse braiding.

We have to stress that all the braiding relations used
in this paper are no more than the cross-multiplication rules for
two copies (primed and unprimed) of the same differential algebra.
Using the Yang-Baxter equation
$$ R R' R = R' R R' \;\; \ \ \ ( R' \equiv \hat{R}_{23})$$
one can show that these rules really define
associative algebras with uniquely ordered
monomials of generators. In other words, all the presented
examples of the braiding transformation $\Psi$ are
proved to be consistent.

\vspace{.3cm}

{\bf 5.} In this paper, we have investigated several examples
of the $GL_{q}(N)$-covariant algebras which are known to be the braided
Hopf algebras. We have shown that the $GL_{q}(N)$-covariant
differential complexes on these algebras admit the braided differential
Hopf algebra structure and the corresponding coproduct is defined
by the formulas (\ref{0}) proposed by Brzezinski \cite{Br} for
the unbraided case. Moreover, it is not hard to demonstrate
that all differential complexes investigated in this paper
are bicovariant with respect to the left and right braided
inner coactions $\Delta_{L}$ and $\Delta_{R}$. All these
observations lead us to the conjecture that the following variant
of the Brzezinski theorem \cite{Br} is valid for the case of
braided Hopf algebras (for the notation see \cite{Br}): \\
{\bf Theorem.} Let $( \Gamma , \; d)$ be a braided bicovariant
differential calculus over a braided Hopf algebra ${\cal B}$.
Then $( \Gamma^{\wedge} , \; d)$ is a differential (exterior)
braided Hopf algebra of ${\cal B}$. Converse statement is also
correct.

We intend to return to the detailed consideration of this
theorem in our next publication.

\vspace{.5cm}

\noindent {\bf Acknowledgments}

\vspace{.3cm}

\noindent
We would like to thank P.N.Pyatov for stimulating discussions.

\vspace{.2cm}

\noindent
This work was supported in part by the Russian Foundation
of Fundamental Research (grant 93-02-3827).

\end{document}